\documentclass[epsfig,12pt]{article}
\usepackage{graphics}
\usepackage{epsfig}
\usepackage{amssymb}
\usepackage{amsmath}
\begin{document}

\baselineskip .7cm

\author{Mai Van Do \thanks{School of Engineering and Applied Sciences, Harvard University, Cambridge, MA 02138.} \ \ 
 Navin Khaneja \thanks{To whom correspondence may be addressed. Email:navinkhaneja@gmail.com} \thanks{Department of Electrical Engineering, IIT Bombay, Powai - 400076, India.}}

\vskip 4em

\title{\bf Broadband Homonuclear Decoupling}

\maketitle 

\begin{center} {\bf Abstract} \end{center}
We present a solution to the problem of decoupling of a homonuclear two-spin system having weak
isotropic scalar coupling. We describe non-selective pulse sequences that create an effective field
perpendicular to the coupling interaction over a broad range of chemical shifts, with a magnitude
proportional to the chemical shifts. Effective decoupling is achieved when the difference in chemical
shifts imprinted on the perpendicular field is sufficiently larger than the coupling between the spins.
The proposed methods scale down the chemical shifts. The pulse sequences may be useful in various
applications in nuclear magnetic resonance spectroscopy.

\vskip 3em

\section{Introduction}

In this paper, we study a classical problem in NMR spectroscopy, the problem of Homonuclear Decoupling. The presence of coupling between nuclear spins alters net magnetic
field seen by elements of a spin pair. In addition to applied magnetic field the spins also see a small field arising due to the magnetic moment of their partners. This field arising due to coupling can be oriented along or against the static field depending on the spin orientation of the magnetic moment. As a result a spin precession frequency of spin $I$ is altered to
$\nu_I +  \frac{J}{2}$ and $\nu_I - \frac{J}{2}$ depending of the orientation of the magnetic moment. This is called splitting of resonance peaks arising due to couplings.
These spin multiplets reduce the signal to noise ratio in NMR experiments and complicate interpretation of the spectrum. It is desirable to find methods by which
couplings between spins can be decoupled and we can observe just the precession of spins in external magnetic field without being effected by couplings.
In this paper, we develop methods to do this spin decoupling using radio frequency irradiation on the spin systems.

\section{Ising Couplings}

To begin with, we consider two homonuclear spins $I$ and $S$ coupled by an Ising coupling. The Hamiltonian of the system takes the form

\begin{equation}
H = \omega_I I_z + \omega_S S_z + 2 \pi J I_z S_z
\end{equation} where $\omega_I$ and $\omega_S$ are chemical shifts of spin $I$ and $S$ and $J$ is the coupling between the spins.
We assume $|\omega_I - \omega_S| \gg J$, so called weakly coupled spin system, as a typical case.   Our goal is to decouple the spins
without precise information of $\omega_I$ and $\omega_s$ and $J$ and yet observe and measure the chemical shifts.

We accomplish this by a a four stage pulse sequence that evolves the following set of Hamiltonians

\begin{eqnarray}
\label{eq:hdbasichamil1}
H_1 &=&  \omega_I I_z + \omega_S S_z + 2 \pi J I_z S_z  + A F_x  \\
H_2 &=&  -\omega_I I_z - \omega_S S_z + 2 \pi J I_z S_z  + A F_x  \\
H_3 &=&  -\omega_I I_z - \omega_S S_z + 2 \pi J I_z S_z  - A F_x  \\
\label{eq:hdbasichamil4} H_3 &=&  \omega_I I_z + \omega_S S_z + 2 \pi J I_z S_z  - A F_x
\end{eqnarray}

We produce the evolution

\begin{equation}
U = \exp(-i H_4 \Delta) \exp(-i H_3 \Delta)\exp(-i H_2 \Delta)\exp(-i H_1 \Delta)
\end{equation}
To understand the effect of this pulse sequence we proceed in the frame of the toggling chemical shift. In this frame the rf-Hamiltonian transforms to

\begin{eqnarray}
AI_x &\rightarrow&  A (I_x \cos \omega_I t  - I_y \sin \omega_I t)  \\
A I_x& \rightarrow&   A (I_x \cos \omega_I (\Delta -t) - I_y \sin \omega_I (\Delta - t))  \\
AI_x &\rightarrow&   -A (I_x \cos \omega_I t  + I_y \sin \omega_I t) \\
AI_x  &\rightarrow&    -A (I_x \cos \omega_I (\Delta -t) + I_y \sin \omega_I (\Delta - t))
\end{eqnarray}
For $\omega_I \Delta < 1$, we can expand  $\sin \omega_I t \sim \omega_I t $. Adding the four evolutions we find everything adds along $y$ direction and we
get

\begin{eqnarray*}
&& \int_0^ {\Delta}  A (I_x \cos \omega_I t  - I_y \sin \omega_I t) dt  +  \int_0^ {\Delta}   A (I_x \cos \omega_I (\Delta -t) - I_y \sin \omega_I (\Delta - t))  \\ &-& \int_0^ {\Delta}  A (I_x \cos \omega_I t  + I_y \sin \omega_I t) - \int_0^ {\Delta}   A (I_x \cos \omega_I (\Delta -t) + I_y \sin \omega_I (\Delta - t)) \\ &=& - 2  A \Delta  \omega_I I_y = - 4 \frac{\theta}{2} \omega_I I_y
\end{eqnarray*} where $\theta = A \Delta$. Therefore in the toggling frame of the chemical shifts, to first order, the net evolution we produce
an effective evolution

\begin{equation}
H_{eff} = -\frac{\theta}{2} ( \omega_I I_y + \omega_S S_y) + 2 \pi J I_zS_z
\end{equation} We have produced an effective rf-field that is perpendicular to the coupling interactions. This will decouple the spins. When we go in the interaction frame of $ ( \omega_I I_y + \omega_S S_y)$, all coupling is eliminated.

\begin{figure}[h]
\begin{center}
\includegraphics[scale=.5]{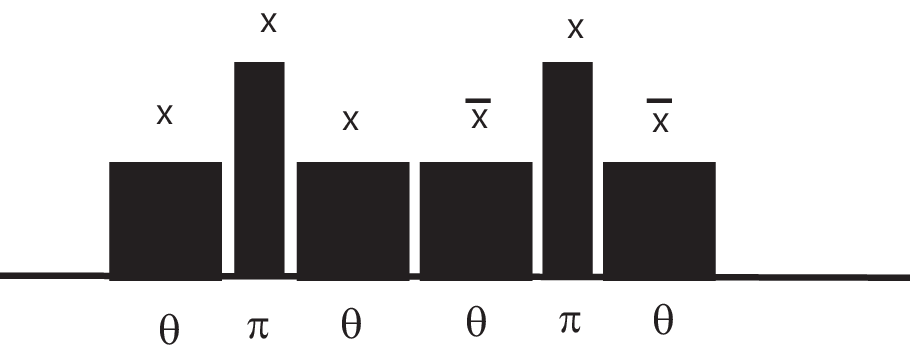}
\end{center}
\caption{The figure shows the pulse sequence for synthesizing an effective $y$ field used for spin decoupling. \label{fig:hdpulsebasic}}
\end{figure}

The pulse sequence for producing the desired effective Hamiltonian is shown in figure. \ref{fig:hdpulsebasic}

\begin{figure}[htb!]
\begin{center}
\includegraphics[scale=.25]{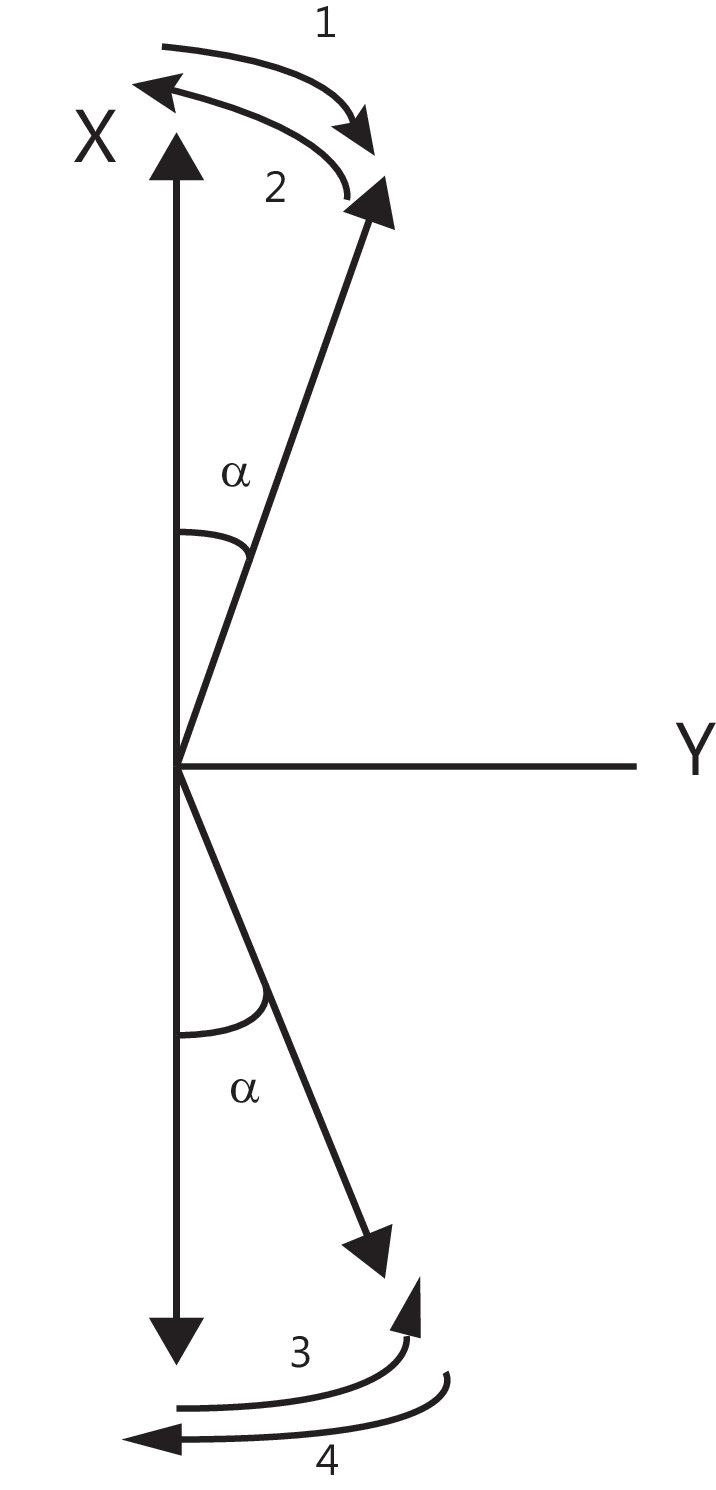}
\end{center}
\label{fig:hdcycle}
\end{figure}

How does the rf Hamiltonian evolve in the toggling frame of the chemical shift is depicted in figure  \ref{fig:hdcycle}.

We apply this building block depicted in figure  \ref{fig:hdpulsebasic} and sample a point from FID (free induction decay) and repeat. The prototype of the basic experiment
is shown in

\begin{figure}[htb!]
\begin{center}
\includegraphics[scale=.5]{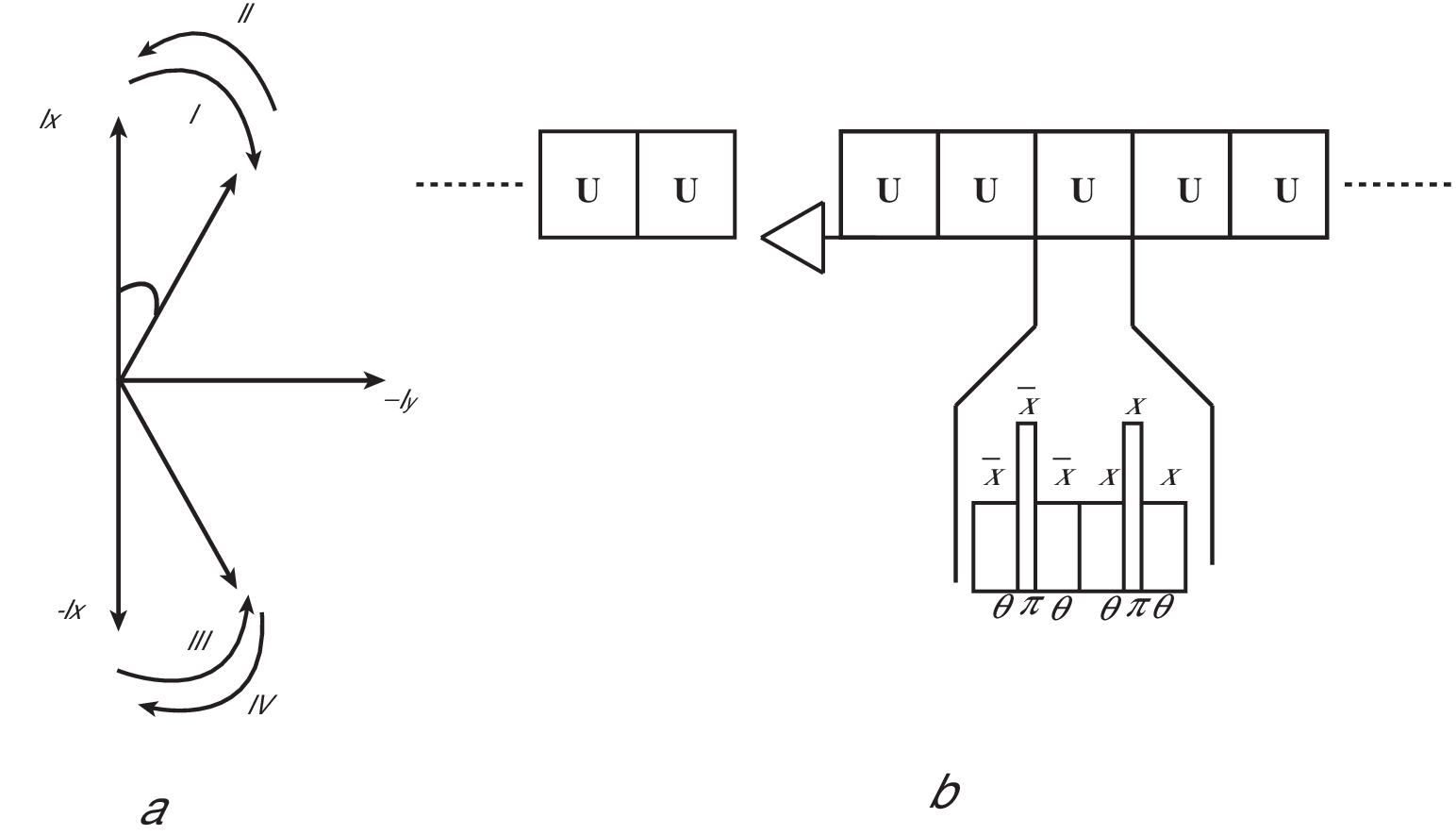}
\end{center}
\label{fig:hdexp}
\end{figure}

We have calculated the effective Hamiltonian to first order in $\theta$. We we calculate it to second order in the theta we find the effective Hamiltonian

\begin{equation}
\exp(-i H_{eff} 4 \Delta ) = \exp(-i H_4 \Delta) \exp(-i H_3 \Delta)\exp(-i H_2 \Delta)\exp(-i H_1 \Delta)
\end{equation} can be calculated using $\exp(A)\exp(B) = \exp(A + B + \frac{[A, B]}{2} + \frac{[A[A, B]] + [[A, B]B]}{12} + \dots )$.

This gives

\begin{equation}
H_{eff} 4 \Delta = -2 \theta ( \omega_I I_y + \omega_S S_y) \Delta + 2 \theta^2  ( \omega_I I_z + \omega_S S_z) \Delta + 2 \pi J \Delta ( 4 I_zS_z + 4 \theta (I_yS_z + I_zS_y)
+ \frac{16}{3} \theta^2 I_yS_y)
\end{equation}

\begin{figure}[htb!]
\begin{center}
\includegraphics[scale=.4]{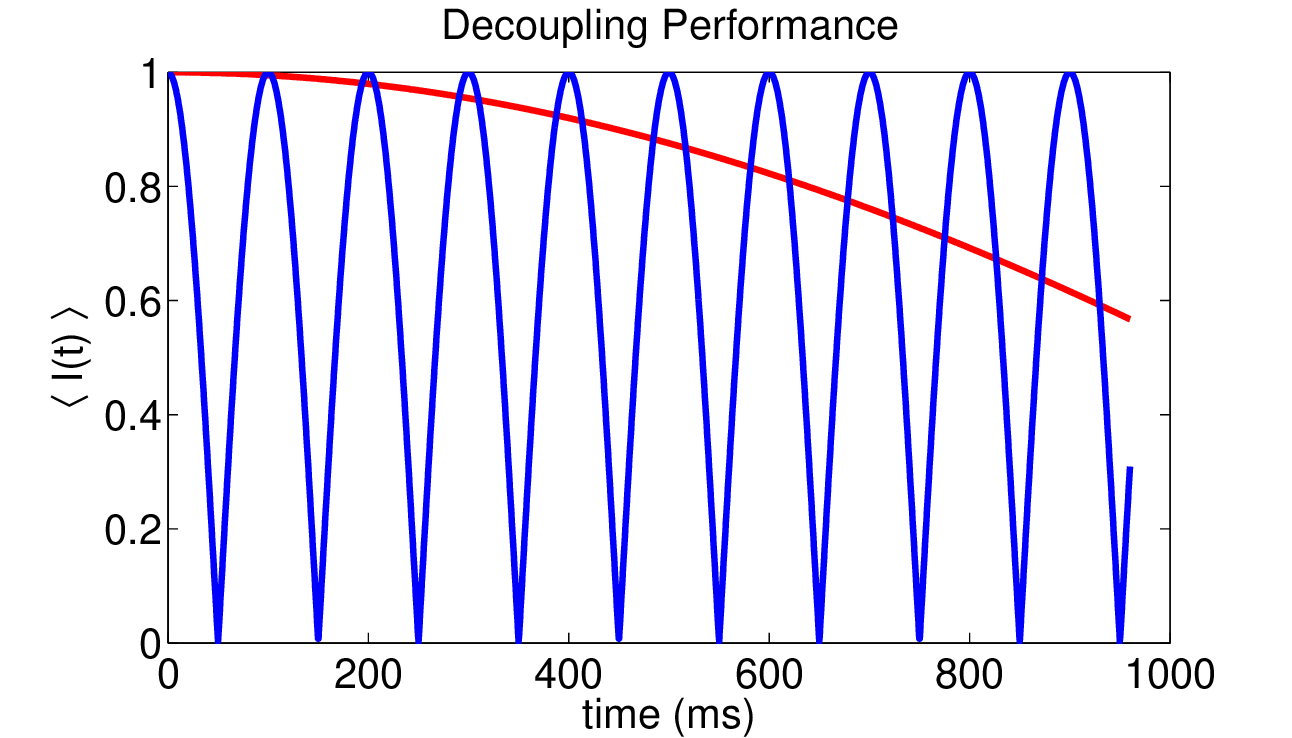}
\includegraphics[scale=.4]{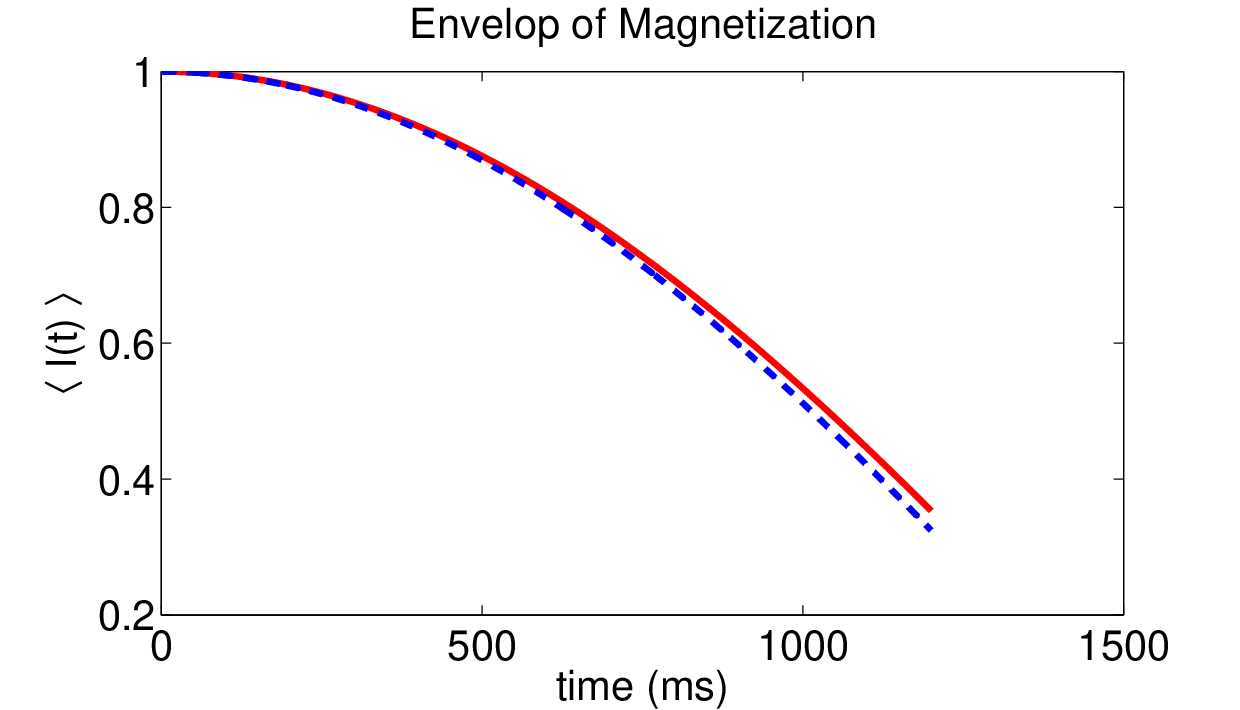}
\end{center}
\caption{The top figure shows how magnetization $\langle I(t) \rangle = \sqrt{\langle I_x(t) \rangle ^2 + \langle I_y(t) \rangle ^2 + \langle I_z(t) \rangle ^2}$ on spin I evolves when we donot decouple (blue curve) vs when we apply the decoupling sequence in Fig. \ref{fig:hdpulsebasic} (red curve). The decoupling of spins results in magnetization of spin $I$ staying on spin $I$. Here, $\frac{\omega_I }{2\pi}= 2400$ Hz, $\frac{\omega_S}{2 \pi} = 800$ Hz, $J= 10$ Hz, $\frac{A }{2 \pi}= 5000$ Hz
and $\theta = \frac{\pi}{10}$. Bottom figure compares the envelop of magnetization of spin $I$ (red curve) with the calculated value as in Eq. \ref{eq:hdeffect} (blue curve).} \label{fig:hdenvising}
\end{figure}

The effective field instead of being along $y$ is along direction $I_{y'}= -I_y + \theta I_z$ and  $ S_{y'} = -S_y + \theta S_z$ respectively. We resolve the coupling along this direction keeping only the parallel
part of the coupling with the effective field. This parallel coupling is

\begin{equation}
4 J_{\parallel} =  2 \pi J \Delta ( 4 \theta^2 - 8 \theta^2 +  \frac{16}{3} \theta^2) I_{y'}S_{y'}
\end{equation} Therefore the effective Hamiltonian is

\begin{equation}
\label{eq:hdeffect}
H_{eff} = \frac{\theta}{2} ( \omega_I I_{y'} + \omega_S S_{y'}) + 2 \pi J  \frac{\theta^2}{3}  I_{y'}S_{y'}
\end{equation} Therefore a part of the coupling stays and reflects itself as an envelop in Fig. \ref{fig:hdenvising}, where
we show how the magnetization on spin $I$ evolves when we donot decouple (blue curve) vs when we apply our decoupling sequence (red curve).
When no decoupling is performed, magnetization evolves to spin $S$ and we see a coupling evolution. When we apply decoupling sequence, the magnetization
on spin $I$ stays

\begin{figure}[htb!]
\begin{center}
\includegraphics[scale=.5]{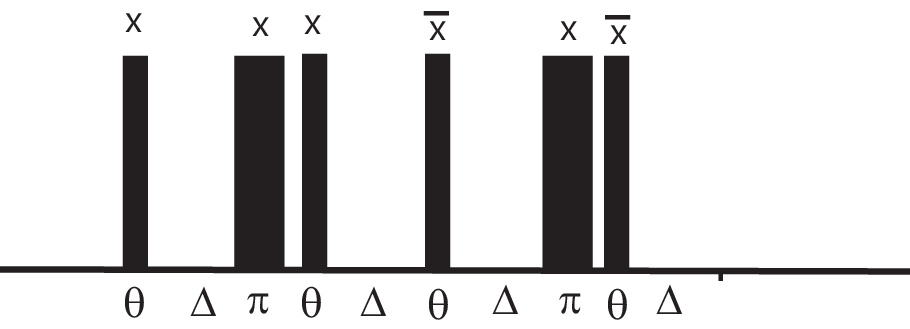}
\end{center}
\caption{The figure shows the pulse sequence for synthesizing an effective $y$ field constructed from pulses and delays.} 
\label{fig:hdpulsebasic1}
\end{figure}

The pulse sequence in Fig. \ref{fig:hdpulsebasic} applies rf-field at the same time as coupling and chemical shift evolution resulting. It is possible to engineer the same effective Hamiltonian
if we apply a hard rf-pulse of flip angle $\theta$ followed by a delay in which coupling and chemical shift evolve. In nutshell the evolution

\begin{equation}
\exp( -i (  \omega_I I_z + \omega_S S_z + 2 \pi J I_z S_z  + A F_x) \Delta) \sim  \exp( -i (  \omega_I I_z + \omega_S S_z + 2 \pi J I_z S_z ) \Delta) \exp( -i  F_x \theta)
\end{equation} is approximated by pulse and evolution as above. We can now write the pulse sequence as

\begin{equation}
U = U_4 U_3 U_2 U_1
\end{equation}

\begin{eqnarray}
U_1 &=& \exp( -i (  \omega_I I_z + \omega_S S_z + 2 \pi J I_z S_z ) \Delta) \exp( -i  F_x \theta)  \\
U_2 &=& \exp( -i ( - \omega_I I_z - \omega_S S_z + 2 \pi J I_z S_z ) \Delta) \exp( -i  F_x \theta)  \\
U_3 &=& \exp( -i (  -\omega_I I_z - \omega_S S_z + 2 \pi J I_z S_z ) \Delta) \exp( i  F_x \theta)  \\
U_4 &=& \exp( -i (  \omega_I I_z + \omega_S S_z + 2 \pi J I_z S_z ) \Delta) \exp( i  F_x \theta)
\end{eqnarray} The pulse sequence for producing the desired effective evolution is given in Fig. \ref{fig:hdpulsebasic1}

We can now calculate the effective Hamiltonian for this pulse sequence.

\begin{equation}
\exp(-i H_{eff} 4 \Delta ) = U_4 U_3 U_2 U_1
\end{equation} can be calculated using $\exp(A)\exp(B) = \exp(A + B + \frac{[A, B]}{2} + \frac{[A[A, B]] + [[A, B]B]}{12} + \dots )$.

This gives

\begin{equation}
H_{eff} 4 \Delta = -2 \theta ( \omega_I I_y + \omega_S S_y) \Delta + 2 \theta^2  ( \omega_I I_z + \omega_S S_z) \Delta + 2 \pi J \Delta ( 4 I_zS_z + 4 \theta (I_yS_z + I_zS_y)
+ \frac{18}{3} \theta^2 I_yS_y )
\end{equation}

\begin{figure}[htb!]
\centering
\includegraphics[scale=.4]{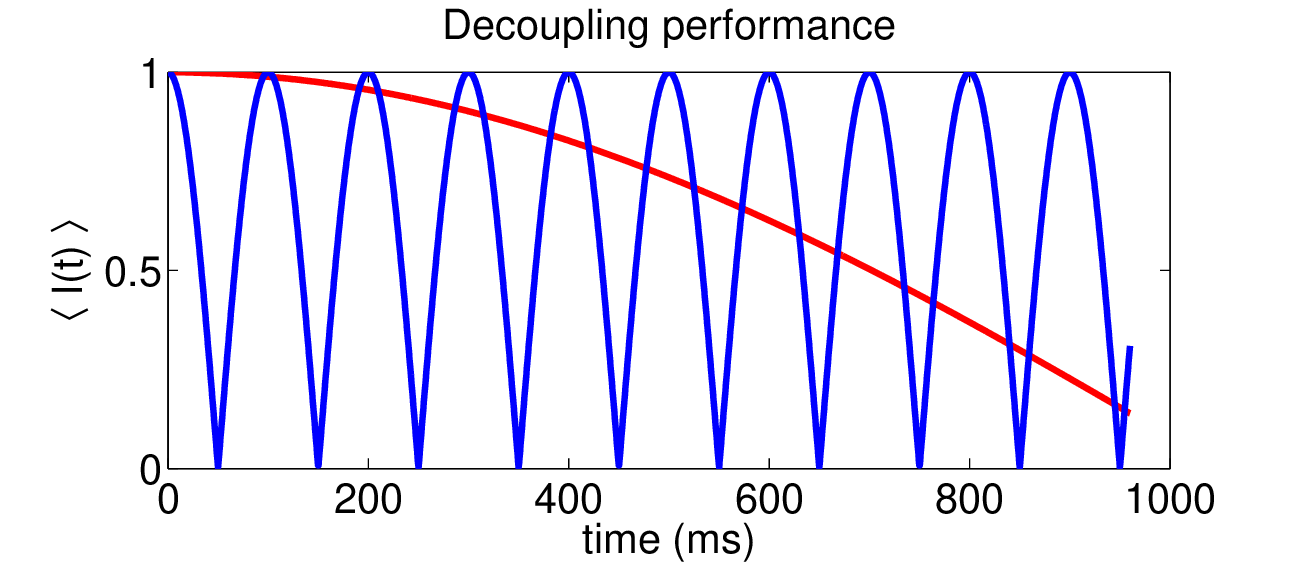}
\includegraphics[scale=.4]{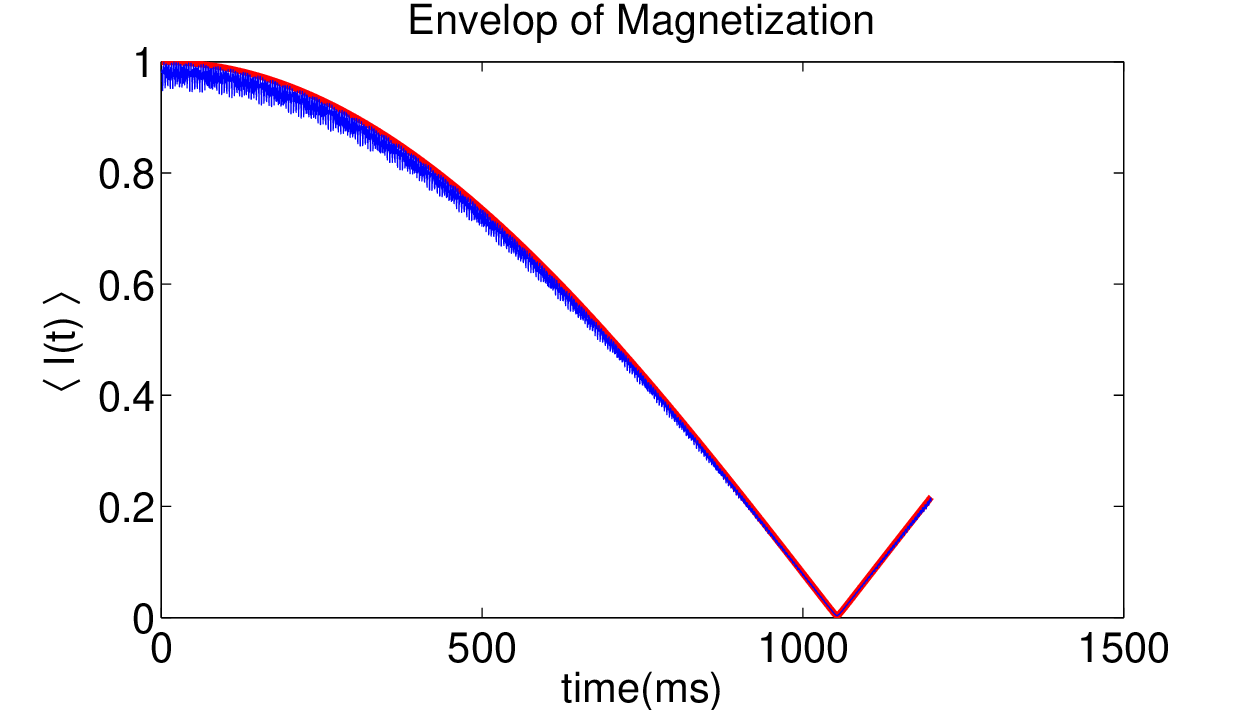}
\caption{The top figure shows how magnetization on spin I evolves when we donot decouple (blue curve) vs when we apply 
the decoupling sequence in Fig. \ref{fig:hdpulsebasic} (red curve). The
decoupling of spins results in magnetization of spin $I$ staying on spin $I$. Here, $\frac{\omega_I }{2\pi}= 2400$ Hz, $\frac{\omega_S}{2 \pi} = 800$ Hz, $J= 10$ Hz, $\frac{A }{2 \pi}= 5000$ Hz
and $\theta = \frac{\pi}{10}$. Bottom figure compares the envelop of magnetization of spin $I$ (red curve) with the calculated value as in Eq. \ref{eq:hdeffect1} (blue curve).} \label{fig:hdenvising1}
\end{figure}

The effective field instead of being along $y$ is along direction $I_{y'}= -I_y + \theta I_z$ and  $ S_{y'} = -S_y + \theta S_z$ respectively. We resolve the coupling along this direction keeping only the parallel
part of the coupling with the effective field. This parallel coupling is

\begin{equation}
4 J_{\parallel} =  2 \pi J \Delta ( 4 \theta^2 - 8 \theta^2 +  \frac{18}{3} \theta^2) I_{y'}S_{y'}
\end{equation} Therefore the effective Hamiltonian is

\begin{equation}
\label{eq:hdeffect1}
H_{eff} = \frac{\theta}{2} ( \omega_I I_{y'} + \omega_S S_{y'}) + 2 \pi J \frac{\theta^2}{2}  I_{y'}S_{y'}
\end{equation} Therefore a part of the coupling stays and reflects itself as an envelop in Fig. \ref{fig:hdenvising1}, where
we show how the magnetization on spin $I$ evolves when we donot decouple (blue curve) vs when we apply our decoupling sequence (red curve).
When no decoupling is performed, magnetization evolves to spin $S$ and we see a coupling evolution. When we apply decoupling sequence, the magnetization
on spin $I$ stays

\begin{figure}[htb!]
\begin{center}
\includegraphics[scale=.5]{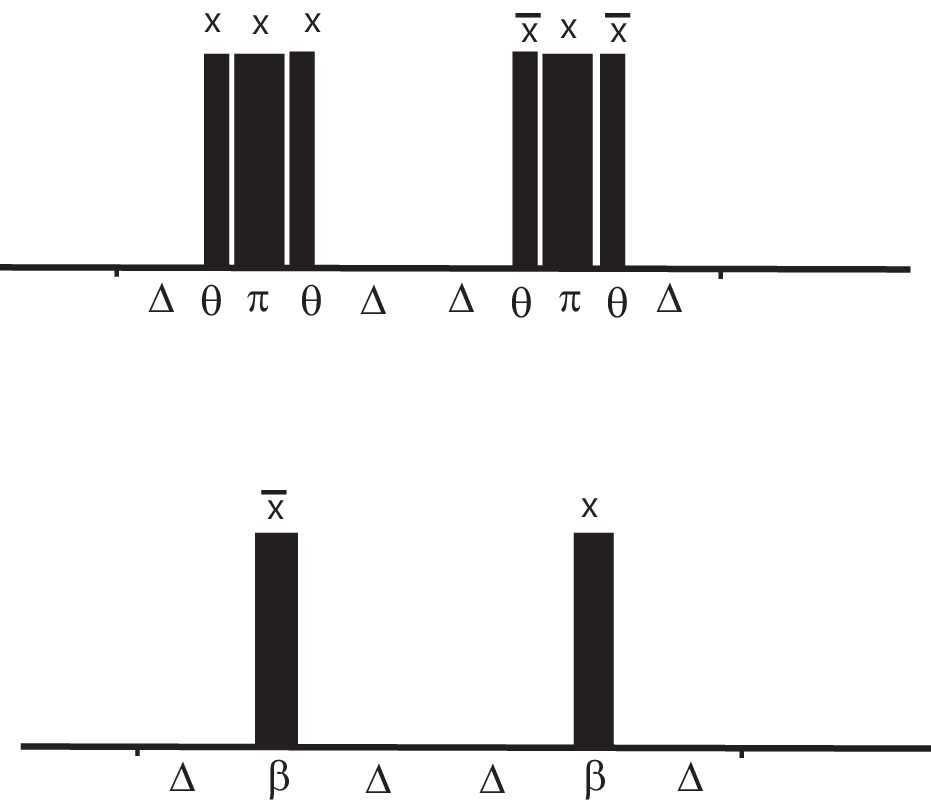}
\end{center}
\caption{The figure shows the pulse sequence for synthesizing an effective $y$ field used for spin decoupling. The pulse sequence enhances the resolution of chemical shifts by a factor of
$2$ over the sequence in \ref{fig:hdpulsebasic}.}\label{fig:hdpulsebasic2}
\end{figure}

In the pulse sequences presented in Fig. \ref{fig:hdpulsebasic} and  Fig. \ref{fig:hdpulsebasic1}, we produce an effective $y$ field with scaling of chemical shift by a factor
$\frac{\theta}{2}$. We now show how by using pulses and delays in our pulses sequence, we can improve this scaling factor to $\theta$. Let

\begin{equation}
U = U_4 U_3 U_2 U_1
\end{equation}

\begin{eqnarray}
U_1 &=&  \exp( -i  F_x \theta) \exp( -i (  \omega_I I_z + \omega_S S_z + 2 \pi J I_z S_z ) \Delta) \\
U_2 &=& \exp( -i ( - \omega_I I_z - \omega_S S_z + 2 \pi J I_z S_z ) \Delta) \exp( -i  F_x \theta)  \\
U_3 &=&  \exp(- i  F_x \theta) \exp( -i (  -\omega_I I_z - \omega_S S_z + 2 \pi J I_z S_z ) \Delta) \\
U_4 &=& \exp( -i (  \omega_I I_z + \omega_S S_z + 2 \pi J I_z S_z ) \Delta) \exp( i  F_x \theta)
\end{eqnarray} The pulse sequence for producing the desired effective evolution is given in Fig. \ref{fig:hdpulsebasic2}

 We can now calculate the effective Hamiltonian for this pulse sequence.

\begin{equation}
\exp(-i H_{eff} 4 \Delta ) = U_4 U_3 U_2 U_1
\end{equation}

\begin{equation}
H_{eff} 4 \Delta = -4 \theta ( \omega_I I_y + \omega_S S_y) \Delta + 4 \theta^2  ( \omega_I I_z + \omega_S S_z) \Delta + 2 \pi J \Delta ( 4 I_zS_z + 4 \theta (I_yS_z + I_zS_y)
+ \frac{24}{3} \theta^2 I_yS_y )
\end{equation}

\begin{figure}[htb!]
\begin{center}
\includegraphics[scale=.4]{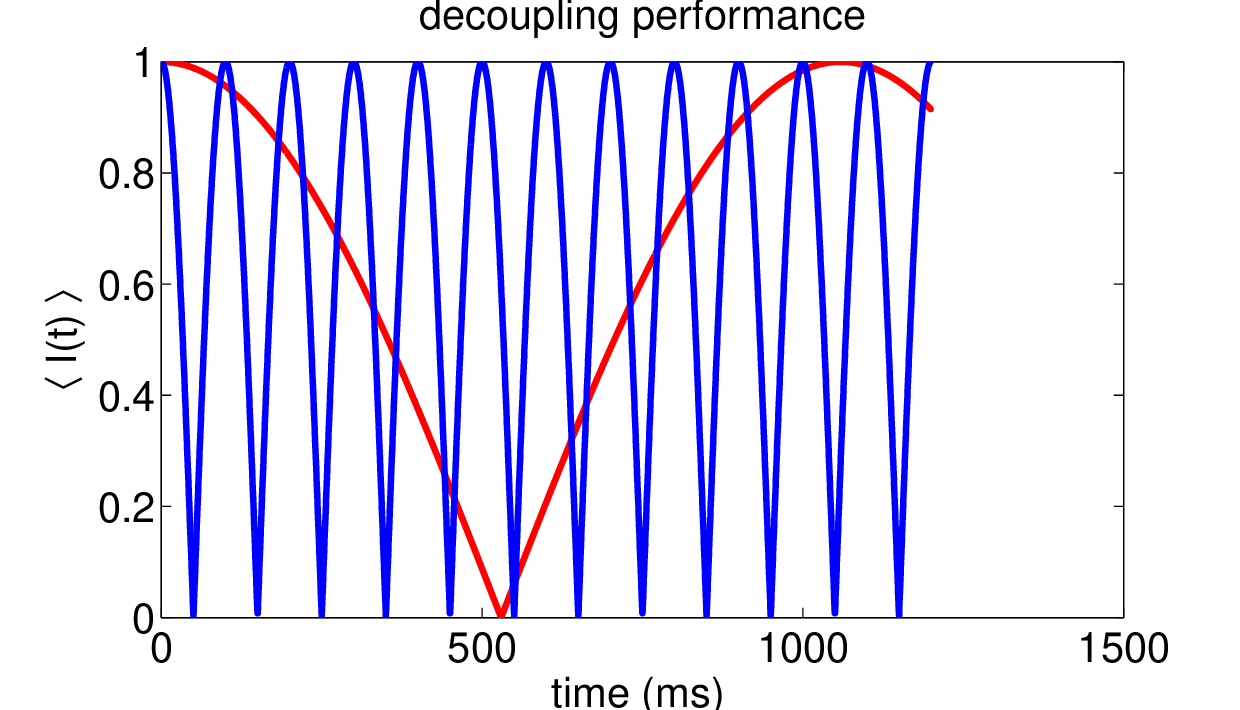}
\includegraphics[scale=.4]{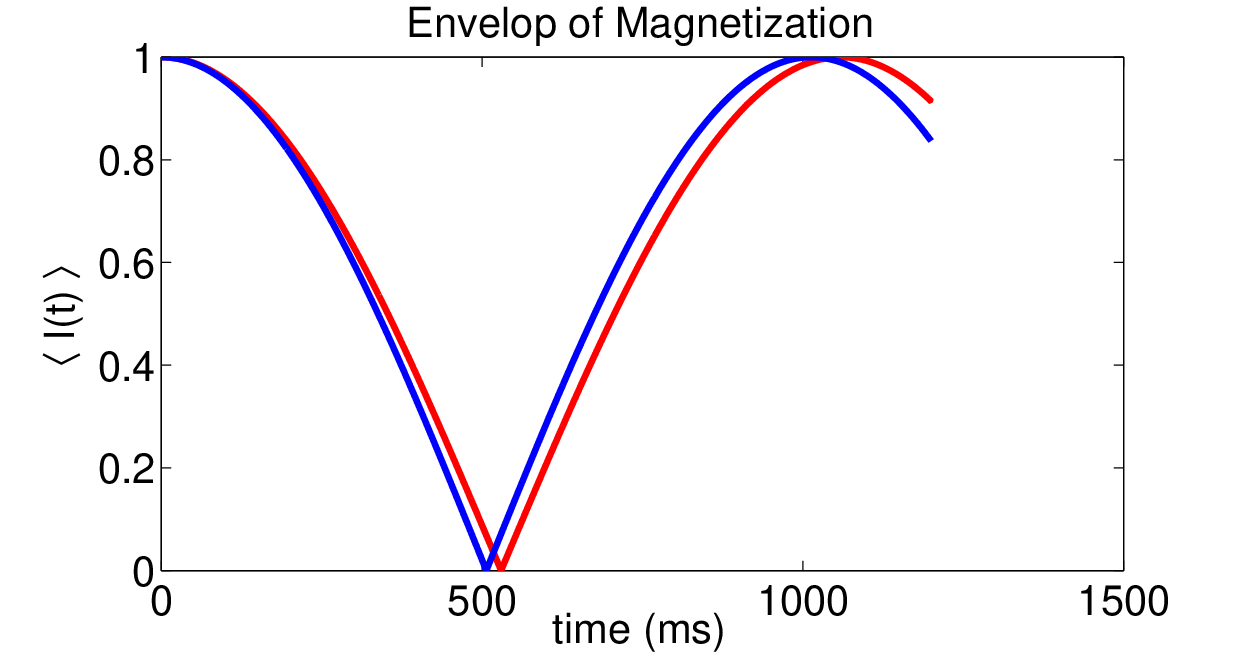}
\end{center}
\caption{The top figure shows how magnetization on spin I evolves when we donot decouple (blue curve) vs when we apply the decoupling sequence in Fig. 
\ref{fig:hdpulsebasic2}(red curve). The
decoupling of spins results in magnetization of spin $I$ staying on spin $I$. Here, $\frac{\omega_I }{2\pi}= 2400$ Hz, $\frac{\omega_S}{2 \pi} = 800$ Hz, $J= 10$ Hz, $\frac{A }{2 \pi}= 5000$ Hz
and $\theta = \frac{\pi}{10}$. Bottom figure compares the envelop of magnetization of spin $I$ (red curve) with the calculated value as in Eq. \ref{eq:hdeffect2} (blue curve).}\label{fig:hdenvising2}
\end{figure}

This parallel coupling is

\begin{equation}
4 J_{\parallel} =  2 \pi J \Delta ( 4 \theta^2 - 8 \theta^2 +  \frac{24}{3} \theta^2) I_{y'}S_{y'}
\end{equation} Therefore the effective Hamiltonian is

\begin{equation}
\label{eq:hdeffect2}
H_{eff} = \theta ( \omega_I I_{y'} + \omega_S S_{y'}) + 2 \pi J \theta^2 I_{y'}S_{y'}
\end{equation} Therefore a part of the coupling stays and reflects itself as an envelop in Fig. \ref{fig:hdenvising2}, where
we show how the magnetization on spin $I$ evolves when we donot decouple (blue curve) vs when we apply our decoupling sequence (red curve).
When no decoupling is performed, magnetization evolves to spin $S$ and we see a coupling evolution. When we apply decoupling sequence, the magnetization
on spin $I$ stays

The pulse sequences we have presented use four decoupling stages for decoupling the Ising coupling. Here we show how we can do this
just using two decoupling stages.

\begin{figure}[h]
\begin{center}
\includegraphics[scale=.5]{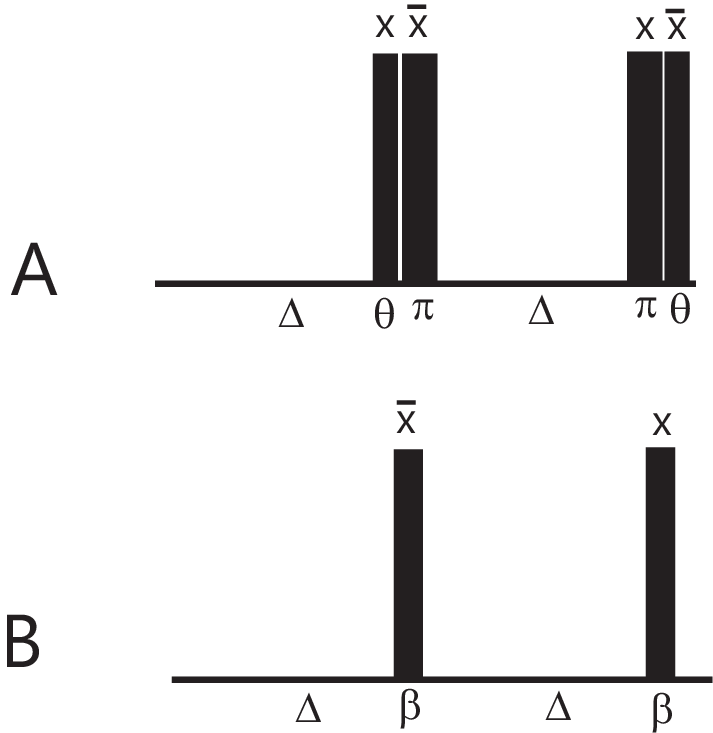}
\end{center}
\caption{The figure shows the pulse sequence for synthesizing an effective $y$ field used for spin decoupling. The pulse sequence uses only two evolution stages.
Pulse B above simplifies Pulse A.} \label{fig:hdpulsebasic3}
\end{figure}

Let

\begin{equation}
U = U_2 U_1
\end{equation}

\begin{eqnarray}
U_1 &=&  \exp( -i (  \omega_I I_z + \omega_S S_z + 2 \pi J I_z S_z ) \Delta) \exp( -i  F_x \theta)\\
U_2 &=& \exp( -i ( - \omega_I I_z - \omega_S S_z + 2 \pi J I_z S_z ) \Delta) \exp( i  F_x \theta)
\end{eqnarray} The pulse sequence for producing the desired effective evolution is given in Fig. \ref{fig:hdpulsebasic3}

 We can now calculate the effective Hamiltonian for this pulse sequence.

\begin{equation}
\exp(-i H_{eff} 4 \Delta ) = U_2 U_1
\end{equation}

\begin{eqnarray}
H_{eff} 2 \Delta &=& -\theta ( \omega_I I_y + \omega_S S_y) \Delta + \frac{\theta}{2} ( \omega_I \theta I_z - \Delta \omega_I^2 I_x) \Delta +  \frac{\theta}{2} ( \omega_S \theta S_z - \Delta \omega_S^2 S_x) \Delta \\
&&2 \pi J \Delta ( 2 I_zS_z + 2 \theta (I_yS_z + I_zS_y) + \theta^2 I_yS_y )
\end{eqnarray}

\begin{figure}[htb!]
\begin{center}
\includegraphics[scale=.4]{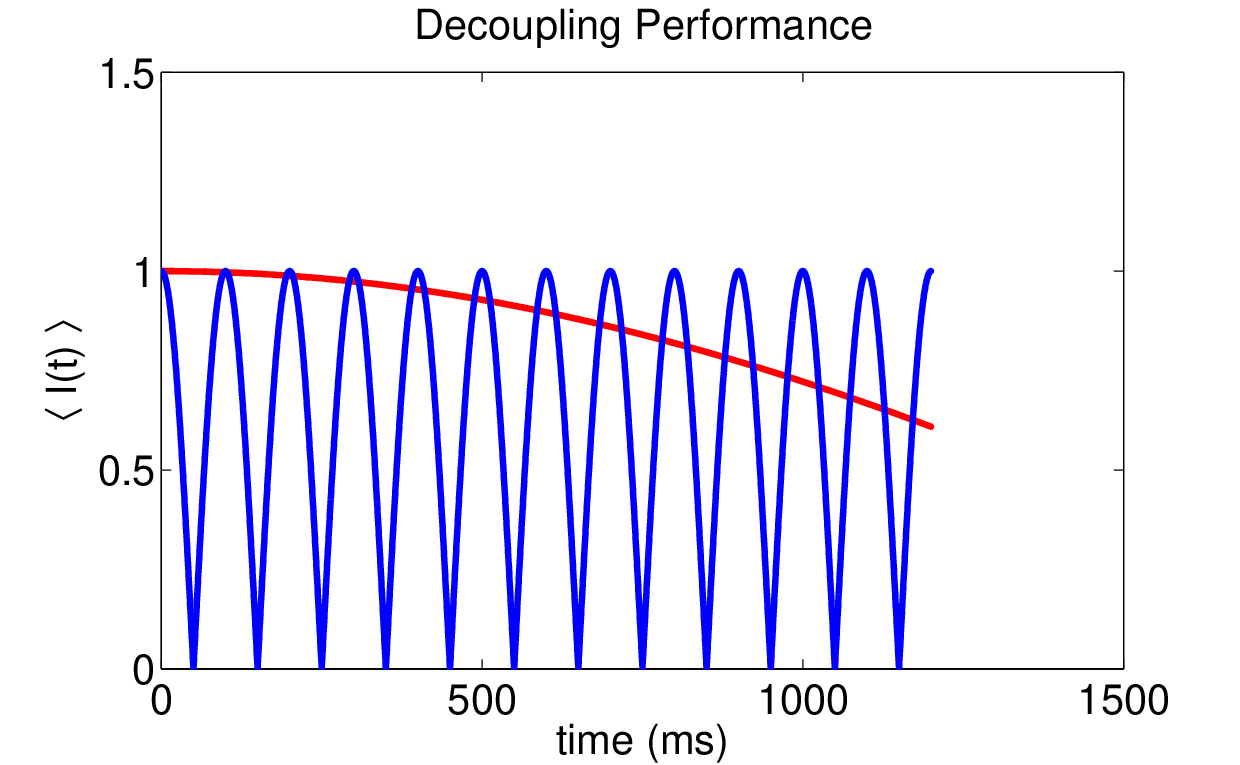}
\includegraphics[scale=.4]{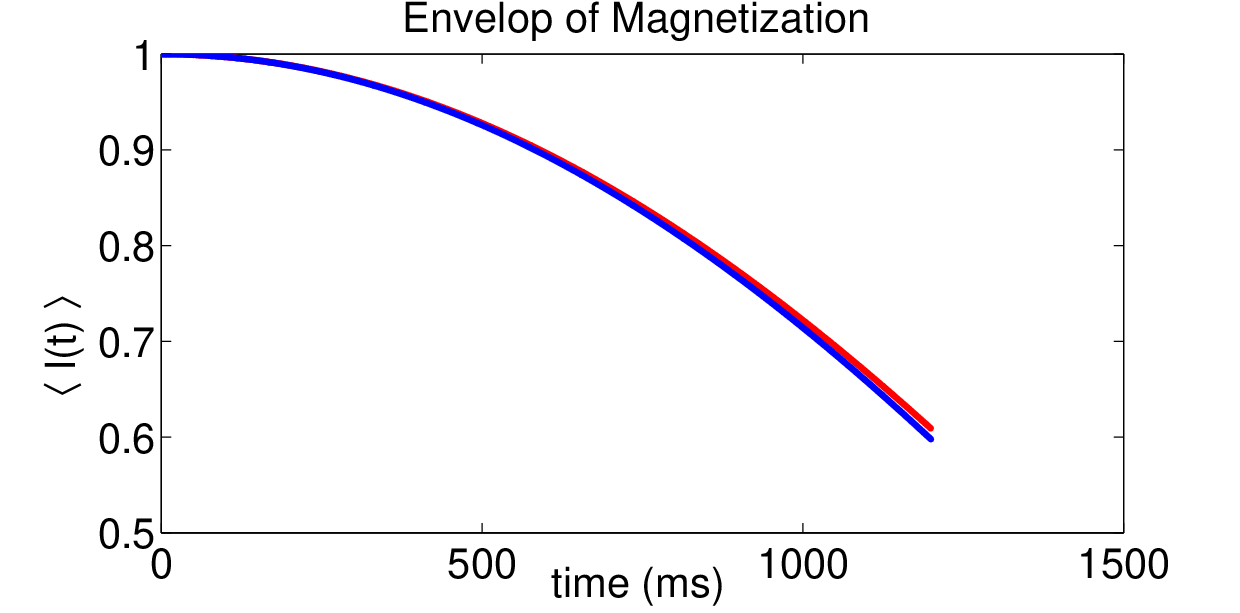}
\end{center}
\caption{The top figure shows how magnetization on spin I evolves when we donot decouple (blue curve) vs when we apply the decoupling sequence in Fig. \ref{fig:hdpulsebasic3}(red curve). The decoupling of spins results in magnetization of spin $I$ staying on spin $I$. Here, $\frac{\omega_I }{2\pi}= 2400$ Hz, $\frac{\omega_S}{2 \pi} = 800$ Hz, $J= 10$ Hz, $\frac{A }{2 \pi}= 5000$ Hz
and $\theta = \frac{\pi}{10}$. Bottom figure compares the envelop of magnetization of spin $I$ (red curve) with the calculated value as in Eq. \ref{eq:hdeffect3} (blue curve)} \label{fig:hdenvising3}
\end{figure}

The effective field is along direction
\begin{eqnarray}
I_\alpha &=& - I_y + \frac{1}{2}(\theta I_z - \omega_I \Delta I_x) \\
S_\beta &=& -S_y + \frac{1}{2}(\theta S_z - \omega_S \Delta S_x)
\end{eqnarray}
This parallel coupling is

\begin{equation}
2 J_{\parallel} =  2 \pi J \Delta ( \frac{\theta^2}{2} - \theta^2 +  \theta^2) I_{\alpha}S_{\beta}
\end{equation} Therefore the effective Hamiltonian is

\begin{equation}
\label{eq:hdeffect3}
H_{eff} = \theta ( \omega_I I_{\alpha} + \omega_S S_{\beta}) + 2 \pi J \frac{\theta^2}{4} I_{\alpha}S_{\beta}
\end{equation} Therefore a part of the coupling stays and reflects itself as an envelop in Fig. \ref{fig:hdenvising3}, where
we show how the magnetization on spin $I$ evolves when we donot decouple (blue curve) vs when we apply our decoupling sequence (red curve).
When no decoupling is performed, magnetization evolves to spin $S$ and we see a coupling evolution. When we apply decoupling sequence, the magnetization
on spin $I$ stays

\section{Isotropic Couplings}
 
\begin{figure}[htb!]
\begin{center}
\begin{tabular}{cc}
\includegraphics[scale=.5]{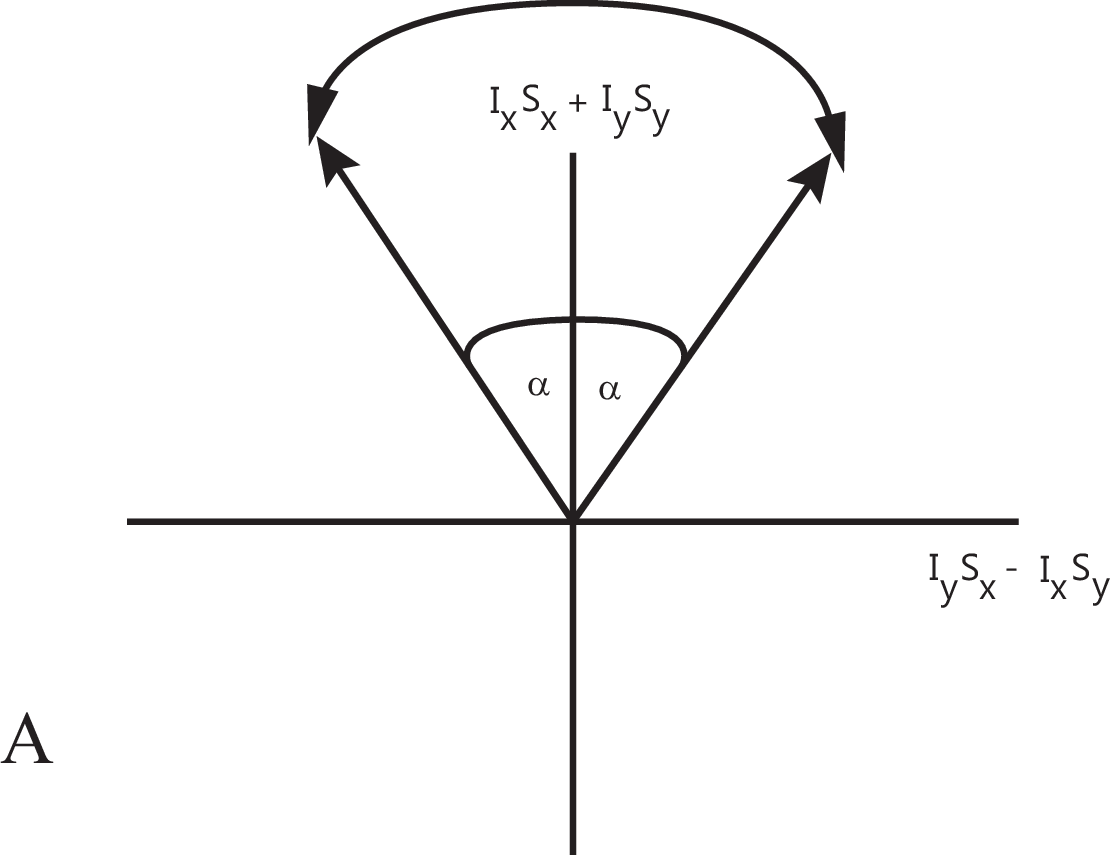} & \includegraphics[scale=.5]{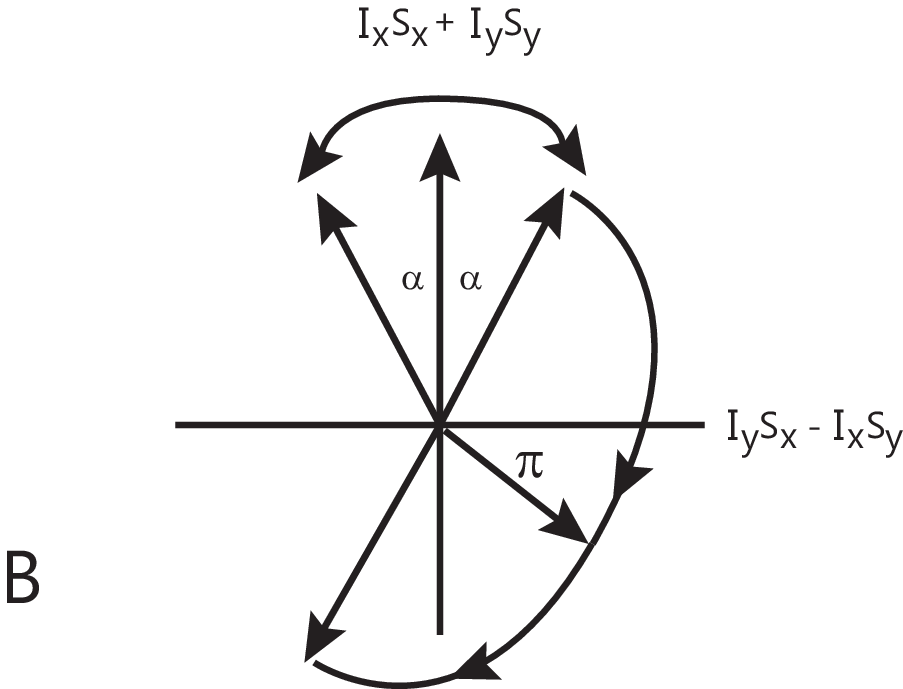}
\end{tabular}
\end{center}
\caption{The top figure shows how the planar part of the Isotropic coupling Hamiltonian evolves under the four cycle of Fig. \ref{fig:hdpulsebasic}.
Bottom figure shows how planar part is averaged by introducing more stages in the pulse sequence shown in figure \ref{fig:hdpulseisobasic}.} \label{fig:planar}
\end{figure}

Until now we considered coupling between the spins as Ising coupling. Now we consider the case of Isotropic coupling, when the coupling Hamiltonian
has the form

\begin{equation}
I \cdot S = I_xS_x + I_yS_y + I_zS_z
\end{equation}
Under the pulse sequence in in Fig. \ref{fig:hdpulsebasic}, we see how this coupling Hamiltonian evolve in the toggling frame of the chemical shifts.
When chemical shifts goes through a evolution $(\omega_I I_z + \omega_S S_z) \Delta $,  $-(\omega_I I_z + \omega_S S_z) \Delta $ ,  $-(\omega_I I_z + \omega_S S_z) \Delta $ ,  $(\omega_I I_z + \omega_S S_z) \Delta $ , the planar part of the coupling $I_xS_x + I_yS_y$ evolves as

 \begin{eqnarray*}
I_xS_x + I_yS_y &\rightarrow&  (I_x S_x + I_yS_y)\cos (\omega_I  - \omega_S) t  +  (I_x S_y - I_yS_x) \sin (\omega_I  - \omega_S) t  = J_1(t) \\
I_xS_x + I_yS_y &\rightarrow&  (I_x S_x + I_yS_y)\cos (\omega_I  - \omega_S) \Delta - t +  (I_x S_y - I_yS_x) \sin (\omega_I  - \omega_S) \Delta - t  = J_2(t) \\
I_xS_x + I_yS_y &\rightarrow&   (I_x S_x + I_yS_y)\cos (\omega_I  - \omega_S) t  -  (I_x S_y - I_yS_x) \sin (\omega_I  - \omega_S) t = J_3(t) \\
I_xS_x + I_yS_y &\rightarrow&   (I_x S_x + I_yS_y)\cos (\omega_I  - \omega_S) \Delta - t -  (I_x S_y - I_yS_x) \sin (\omega_I  - \omega_S) \Delta - t = J_4(t)
\end{eqnarray*} All of this then sums to an effective evolution $ \int_0^\Delta J_1(t) + J_2(t) + J_3(t) + J_4(t) dt $,

\begin{equation}
\label{eq:effj}
\frac{4 (I_x S_x + I_yS_y)\sin (\omega_I  - \omega_S) \Delta }{\omega_I  - \omega_S} \\
\end{equation} The effective rf-field produced by the pulse sequence in Fig. \ref{fig:hdpulsebasic},  is along $y$ direction. This effective field will not decouple the
$yy$ part of the Hamiltonian in \ref{eq:effj}. Therefore we donot get complete decoupling. This is shown in top figure \ref{fig:planar}.

\begin{figure}[h]
\begin{center}
\includegraphics[scale=.5]{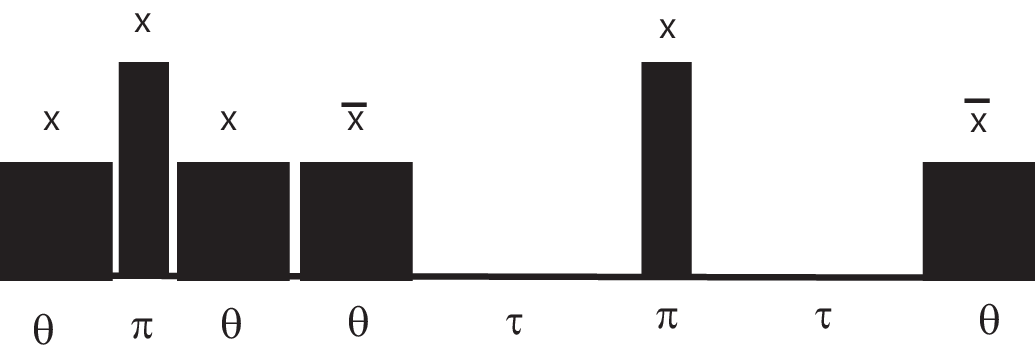}
\end{center}
\caption{The top figure is a six stage pulse sequence to decouple Isotropic coupling Hamiltonian. \label{fig:hdpulseisobasic}}
\end{figure}

To eliminate the $yy$ part of the Hamiltonian we propose the following pulse sequence in Fig. \ref{fig:hdpulseisobasic}, where
$\tau = \frac{\pi}{\omega_I - \omega_S}$ and $\alpha = (\omega_I - \omega_S)\Delta$.

We analyze the six stage pulse sequence formed out of Hamiltonians. The evolution of the planar Hamiltonian under these set of Hamiltonians
is shown in bottom figure \ref{fig:planar}.

\begin{eqnarray}
H_1 &=&  \omega_I I_z + \omega_S S_z + 2 \pi J I \cdot S  + A F_x  \\
H_2 &=&  -\omega_I I_z - \omega_S S_z + 2 \pi J I \cdot S  + A F_x  \\
H_3 &=&  -\omega_I I_z - \omega_S S_z + 2 \pi J I \cdot S - A F_x  \\
H_4 &=&  -\omega_I I_z - \omega_S S_z + 2 \pi J I \cdot S \\
H_5 &=&  \omega_I I_z  + \omega_S S_z + 2 \pi J I \cdot S  \\
H_6 &=&  \omega_I I_z + \omega_S S_z + 2 \pi J I \cdot S  - A F_x
\end{eqnarray} Under this pulse sequence , to first order the planar Hamiltonian averages to

\begin{equation}
\label{eq:effj1}
\frac{4 (I_y S_x- I_xS_y)\cos (\omega_I  - \omega_S) \Delta }{\omega_I  - \omega_S} \\
\end{equation} and the effective rf field Hamiltonian to first order in $\theta$ is as in \ref{fig:hdpulsebasic} $ \frac{\theta}{2}(\omega_I I_y + \omega_S S_y)$.
This effective rf-Hamiltonian will remove the averaged Isotropic coupling and we get decoupling. We can evaluate the effect of the pulse sequence
in  \ref{fig:hdpulseisobasic} to higher order in $\theta$

\begin{equation}
U =  \exp(-i H_6 \Delta) \exp(-i H_5 \tau) \exp(-i H_4 \tau) \exp(-i H_3 \Delta)\exp(-i H_2 \Delta)\exp(-i H_1 \Delta).
\end{equation}

The effective Hamiltonian can now be evaluated. First observe , by evaluating in the toggling frame of the chemical shift we find

$$  \exp(-i H_5 \tau) \exp(-i H_4 \tau) = \exp(-i \frac{4 \pi J \Delta}{\alpha} (2I_yS_x -2 I_xS_y + \pi I_zS_z)) = \exp(-i J_I \Delta) $$

Then the evolution

\begin{eqnarray}
\label{Eq: U6stage}
U &=&  \exp(-i H_6 \Delta) \exp(-i J_I \Delta ) \exp(-i H_3 \Delta)\exp(-i H_2 \Delta)\exp(-i H_1 \Delta) \\ &=& \exp(-i \tilde{J}_I \Delta )  \exp(-i H_6 \Delta)  \exp(-i H_3 \Delta)\exp(-i H_2 \Delta)\exp(-i H_1 \Delta)
\end{eqnarray}where

\begin{equation}
\tilde{J}_I =  \exp(-i H_6 \Delta) J_I  \exp(i H_6 \Delta)
\end{equation} Under this transformation

\begin{eqnarray}
\label{eq:JItransform}
I_yS_x - I_xS_y &\rightarrow&  I_yS_x - I_xS_y  -  \alpha (I_xS_x +  I_yS_y)  \\
I_zS_z &\rightarrow&   I_zS_z + (I_zS_y + I_yS_z) \theta
\end{eqnarray}

In Eq. \ref{Eq: U6stage}, we have already solved for the evolution

\begin{equation}
\exp(-i H_{eff} 4 \Delta ) = \exp(-i H_6 \Delta)  \exp(-i H_3 \Delta)\exp(-i H_2 \Delta)\exp(-i H_1 \Delta)
\end{equation} in Eq. \ref{eq:hdeffect}. The result can be modified for isotropic coupling to

\begin{equation}
\label{eq:isoeffective4}
 H_{eff} = \frac{\theta}{2} ( \omega_I I_{y'} + \omega_S S_{y'}) + 2 \pi J  ( I \cdot S - \frac{\alpha^2}{6}  (I_{x}S_x + I_yS_y ))
\end{equation}

\begin{figure}[htb!]
\begin{center}
\includegraphics[scale=.5]{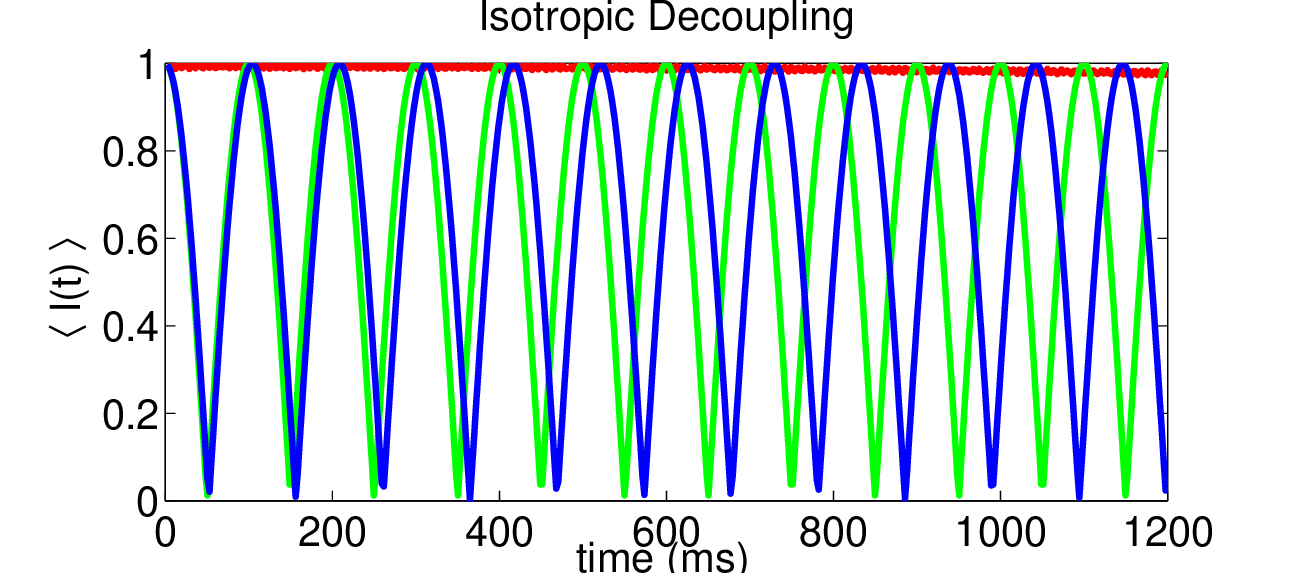}
\end{center}
\caption{ The top figure shows how magnetization on spin I evolves when we donot decouple (green curve) vs when we apply the four stage decoupling sequence in Fig. \ref{fig:hdpulsebasic}(blue curve) and finally the six stage pulse sequence (red curve) of Fig. \ref{fig:hdpulseisobasic}. The decoupling of spins results in magnetization of spin $I$ staying on spin $I$. Here, $\frac{\omega_I }{2\pi}= 2400$ Hz, $\frac{\omega_S}{2 \pi} = 800$ Hz, $J= 10$ Hz, $\frac{A }{2 \pi}= 1000$ Hz and $\theta = \frac{\pi}{10}$} \label{fig:hdenviso}
\end{figure}

Combining eq. \ref{Eq: U6stage} with \ref{eq:JItransform} and \ref{eq:isoeffective4}, we are left with an effective coupling
which is

\begin{equation}
\label{eq:hdeffectiso}
H_{eff} = \frac{\theta}{2} ( \omega_I I_{y'} + \omega_S S_{y'}) - 2 \pi J (\frac{\theta^2}{2} + \frac{\alpha^2}{6}) I_{y'}S_{y'}
\end{equation} Therefore a part of the coupling stays and reflects itself as an envelop in Fig. \ref{fig:hdenviso}, where
we show how the magnetization on spin $I$ evolves when we donot decouple (blue curve) vs when we apply our decoupling sequence (red curve).
When no decoupling is performed, magnetization evolves to spin $S$ and we see a coupling evolution. When we apply decoupling sequence, the magnetization
on spin $I$ stays

When we studied the four stage pulse sequence in Fig. \ref{fig:hdpulsebasic}, we showed it can be implemented in two other ways presented in
Fig.  \ref{fig:hdpulsebasic1} and \ref{fig:hdpulsebasic2}, where we use pulses and delays. For decoupling isotropic interactions, the six stage pulse
sequence of Fig. \ref{fig:hdpulseisobasic} can also be implemented in two other ways. We present there other methods now.

Consider the pulse sequence in the following Fig. \ref{fig:hdpulseisobasic1}, where we have six stages in the evolution.

\begin{figure}[htb!]
\begin{center}
\includegraphics[scale=.5]{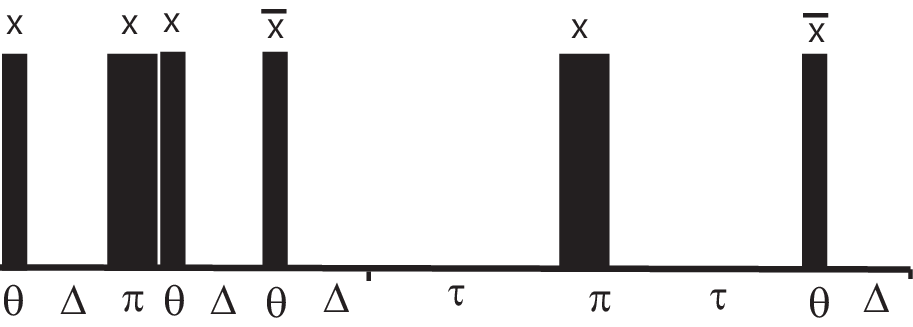}
\end{center}
\caption{The top figure is a six stage pulse sequence to decouple Isotropic coupling Hamiltonian.} \label{fig:hdpulseisobasic1}
\end{figure}

We analyze the six stage pulse sequence formed out of Hamiltonians. Here $\tau = \frac{1}{2 (\omega_I - \omega_S)}$.

\begin{eqnarray}
U_1 &=&  \exp(-i \Delta (\omega_I I_z + \omega_S S_z + 2 \pi J I \cdot S)) \exp(-i \Delta A F_x)  \\
U_2 &=&  \exp(-i \Delta (-\omega_I I_z - \omega_S S_z + 2 \pi J I \cdot S)) \exp(-i \Delta A F_x)  \\
U_1 &=&  \exp(-i \Delta (-\omega_I I_z - \omega_S S_z + 2 \pi J I \cdot S)) \exp(i \Delta A F_x)  \\
U_1 &=&  \exp(-i \tau (-\omega_I I_z - \omega_S S_z + 2 \pi J I \cdot S)) \\
U_1 &=&  \exp(-i \tau (\omega_I I_z + \omega_S S_z + 2 \pi J I \cdot S))  \\
U_1 &=&  \exp(-i \Delta (\omega_I I_z + \omega_S S_z + 2 \pi J I \cdot S)) \exp(i \Delta A F_x)
\end{eqnarray} The total evolution is

\begin{equation}
U = U_6 U_5 U_4 U_3 U_2 U_1
\end{equation}

\begin{figure}[htb!]
\begin{center}
\includegraphics[scale=.5]{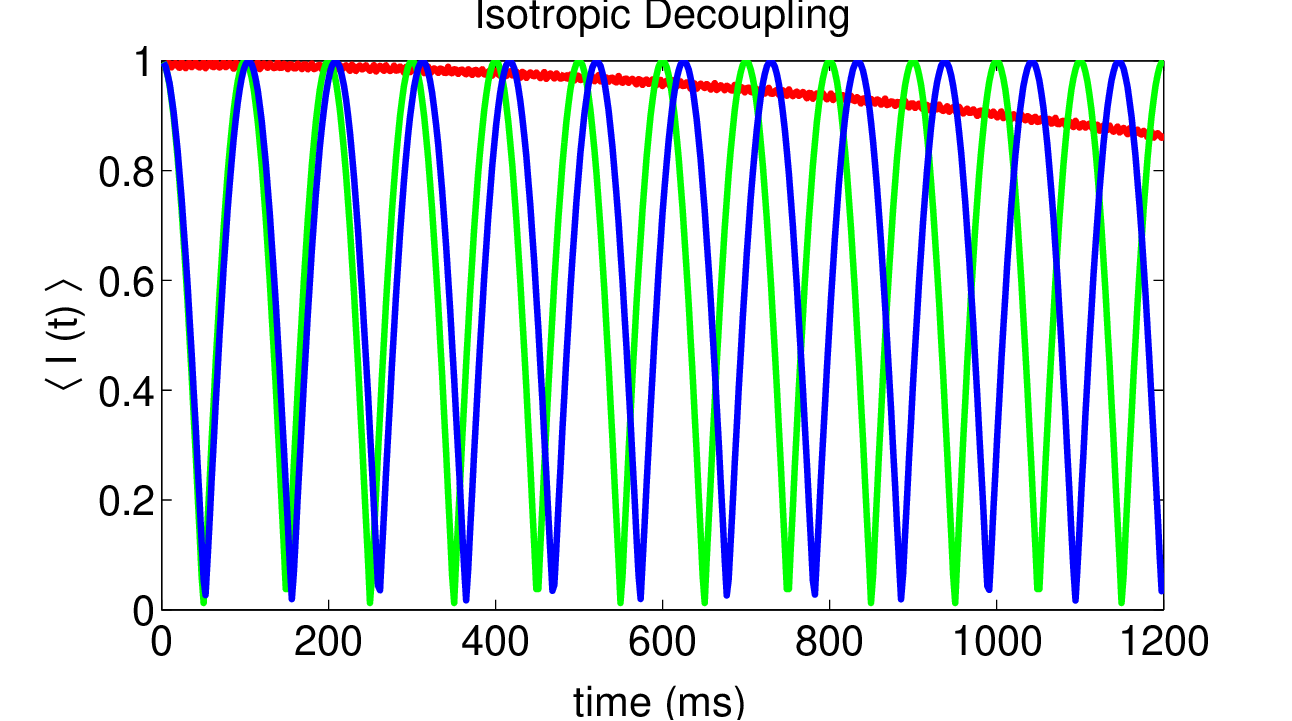}
\end{center}
\caption{ The top figure shows how magnetization on spin I evolves when we donot decouple (green curve) vs when we apply the four stage decoupling sequence in Fig. \ref{fig:hdpulsebasic}(blue curve) and finally the six stage pulse sequence (red curve) of Fig. \ref{fig:hdpulseisobasic1}. The decoupling of spins results in magnetization of spin $I$ staying on spin $I$. Here, $\frac{\omega_I }{2\pi}= 2400$ Hz, $\frac{\omega_S}{2 \pi} = 800$ Hz, $J= 10$ Hz, $\frac{A }{2 \pi}= 1000$ Hz and $\theta = \frac{\pi}{10}$} \label{fig:hdenviso1}
\end{figure}

The effective Hamiltonian of this pulse sequence is

\begin{equation}
\label{eq:hdeffectiso1}
H_{eff} = \frac{\theta}{2} ( \omega_I I_{y'} + \omega_S S_{y'}) - 2 \pi J (\frac{\alpha^2}{6}) I_{y'}S_{y'}
\end{equation} Therefore a part of the coupling stays and reflects itself as an envelop in Fig. \ref{fig:hdenviso1}, where
we show how the magnetization on spin $I$ evolves when we donot decouple (blue curve) vs when we apply our decoupling sequence (red curve).
When no decoupling is performed, magnetization evolves to spin $S$ and we see a coupling evolution. When we apply decoupling sequence, the magnetization
on spin $I$ stays

Now we can again improve the scaling factor of the chemical shifts by changing the order of pulses and delays. Consider the pulse sequence in the following Fig. \ref{fig:hdpulseisobasic2}, where we have six stages in the evolution.

\begin{figure}[htb!]
\begin{center}
\includegraphics[scale=.5]{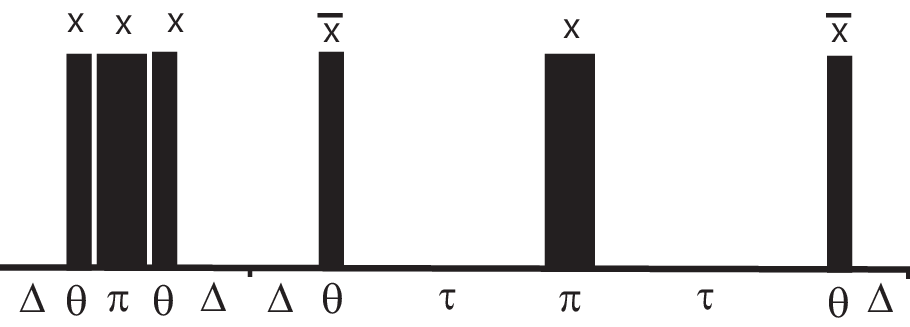}
\end{center}
\caption{The top figure is a six stage pulse sequence to decouple Isotropic coupling Hamiltonian.} \label{fig:hdpulseisobasic2}
\end{figure}

We analyze the six stage pulse sequence formed out of Hamiltonians. Here $\tau = \frac{1}{2 (\omega_I - \omega_S)}$.

\begin{eqnarray}
U_1 &=&  \exp(-i \Delta A F_x)  \exp(-i \Delta (\omega_I I_z + \omega_S S_z + 2 \pi J I \cdot S)) \\
U_2 &=&  \exp(-i \Delta (-\omega_I I_z - \omega_S S_z + 2 \pi J I \cdot S)) \exp(-i \Delta A F_x)  \\
U_1 &=&  \exp(i \Delta A F_x)  \exp(-i \Delta (-\omega_I I_z - \omega_S S_z + 2 \pi J I \cdot S))   \\
U_1 &=&  \exp(-i \tau (-\omega_I I_z - \omega_S S_z + 2 \pi J I \cdot S)) \\
U_1 &=&  \exp(-i \tau (\omega_I I_z + \omega_S S_z + 2 \pi J I \cdot S))  \\
U_1 &=&  \exp(-i \Delta (\omega_I I_z + \omega_S S_z + 2 \pi J I \cdot S)) \exp(i \Delta A F_x)
\end{eqnarray} The total evolution is

\begin{equation}
U = U_6 U_5 U_4 U_3 U_2 U_1
\end{equation}

The effective Hamiltonian of this pulse sequence is

\begin{equation}
\label{eq:hdeffectiso2}
H_{eff} = \theta ( \omega_I I_{y'} + \omega_S S_{y'}) - 2 \pi J (\frac{\alpha^2}{6}) I_{y'}S_{y'}
\end{equation} Therefore a part of the coupling stays and reflects itself as an envelop in Fig. \ref{fig:hdenviso1}, where
we show how the magnetization on spin $I$ evolves when we donot decouple (blue curve) vs when we apply our decoupling sequence (red curve).
When no decoupling is performed, magnetization evolves to spin $S$ and we see a coupling evolution. When we apply decoupling sequence, the magnetization
on spin $I$ stays

\begin{figure}[htb!]
\begin{center}
\includegraphics[scale=.5]{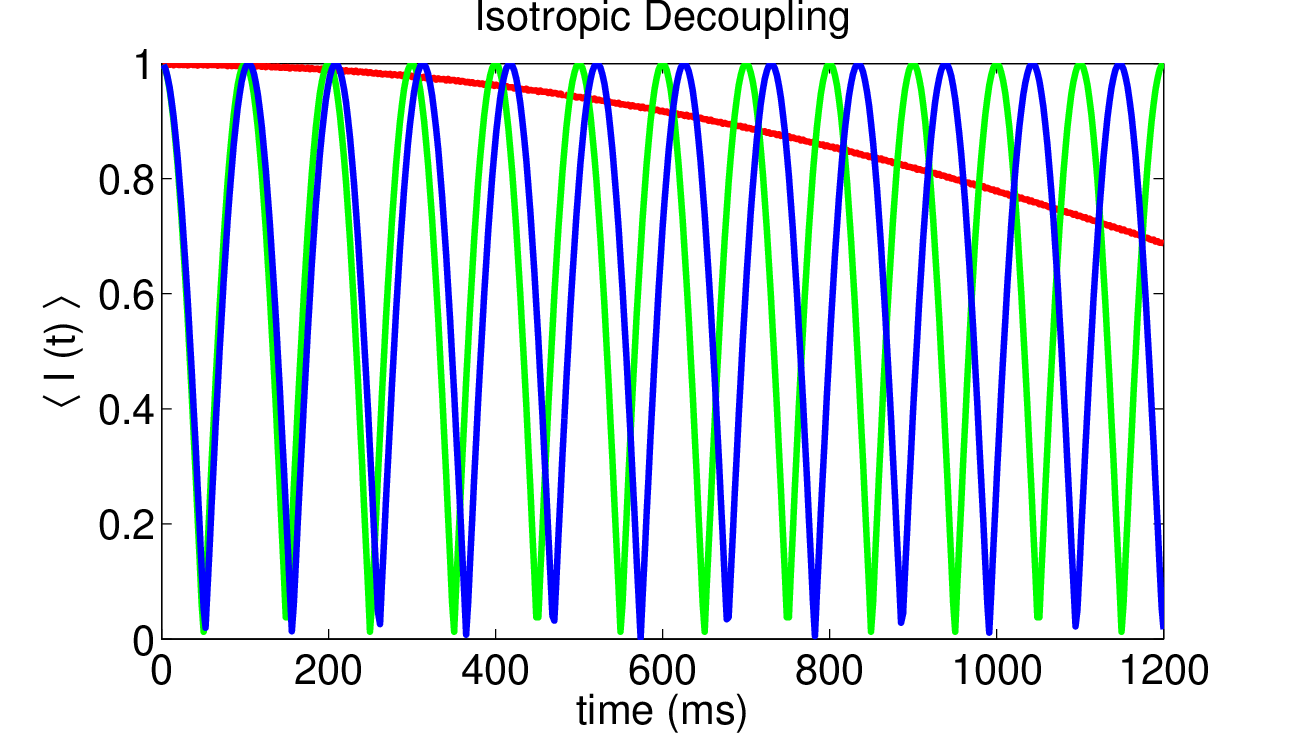}
\end{center}
\caption{ The top figure shows how magnetization on spin I evolves when we donot decouple (green curve) vs when we apply the four stage decoupling sequence in Fig. \ref{fig:hdpulsebasic}(blue curve) and finally the six stage pulse sequence (red curve) of Fig. \ref{fig:hdpulseisobasic2}. The decoupling of spins results in magnetization of spin $I$ staying on spin $I$. Here, $\frac{\omega_I }{2\pi}= 2400$ Hz, $\frac{\omega_S}{2 \pi} = 800$ Hz, $J= 10$ Hz, $\frac{A }{2 \pi}= 1000$ Hz and $\theta = \frac{\pi}{10}$} \label{fig:hdenviso2}
\end{figure}

Until now we have used a six stage pulse sequence to decouple Isotropic coupling. We now show how the four stage pulse sequence in Fig. 
\ref{fig:hdpulsebasic} can be used to decouple isotropic coupling. In the four stage pulse sequence  in Fig. 
\ref{fig:hdpulsebasic}, we choose the time $\Delta = \frac{\pi}{\omega_I - \omega S}$. Since the evolution time $\Delta$ is made large, in the toggling frame of the
chemical shifts we make large excursion/dip. We consider the evolution of the Hamiltonians in Eq. \ref{eq:hdbasichamil1}-\ref{eq:hdbasichamil4}, in the toggling frame
of the chemical shifts. Let $\tau = \Delta - t$

\begin{eqnarray}
\label{eq:largedip}
\tilde H_1(t)  &=&  \exp(i(\omega_I I_z + \omega_S S_z)t) (2 \pi J I \cdot S  + A F_x)\exp(-i(\omega_I I_z + \omega_S S_z)t) \\
\tilde H_2(t)  &=&  \exp(-i(\omega_I I_z + \omega_S S_z) \tau ) (2 \pi J I \cdot S  + A F_x)\exp(i(\omega_I I_z + \omega_S S_z)\tau) \\
\tilde H_3(t)  &=&  \exp(-i(\omega_I I_z + \omega_S S_z)t) (2 \pi J I \cdot S  - A F_x)\exp(i(\omega_I I_z + \omega_S S_z)t) \\
\tilde H_4(t)  &=&  \exp(i(\omega_I I_z + \omega_S S_z)\tau) (2 \pi J I \cdot S  - A F_x) \exp(-i(\omega_I I_z + \omega_S S_z)\tau) 
\end{eqnarray}

We produce the evolution

\begin{equation}
\label{eq:efflargedip}
\exp(-i H_{eff} 4 \Delta) = U = \tilde U_4 \tilde U_3 \tilde U_2 \tilde U_1
\end{equation}where, $\tilde U_i = \exp(A_i + B_i + C_i)$

\begin{eqnarray}
A_i &=& -i \int_0^{\Delta} \tilde H_i(\sigma_1) d \sigma_1 \\
B_i &=& - \int_0^{\Delta} \int_0^{\sigma_1} [\tilde H_i(\sigma_1), \tilde H_i(\sigma_2)] d \sigma_2 d \sigma_1 \\
\nonumber C_i &=&  i  \int_0^{\Delta} \int_0^{\sigma_1} \int_0^{\sigma_2} [\tilde H_i (\sigma_1) [\tilde H_i(\sigma_2), \tilde H_i(\sigma_3)]] +  [ [\tilde H_i (\sigma_1), \tilde H_i(\sigma_2)] \tilde H_i(\sigma_3)]d \sigma_3 d \sigma_2 d \sigma_1 
\end{eqnarray}

\begin{figure}[h]
\begin{center}
\includegraphics[scale=.5]{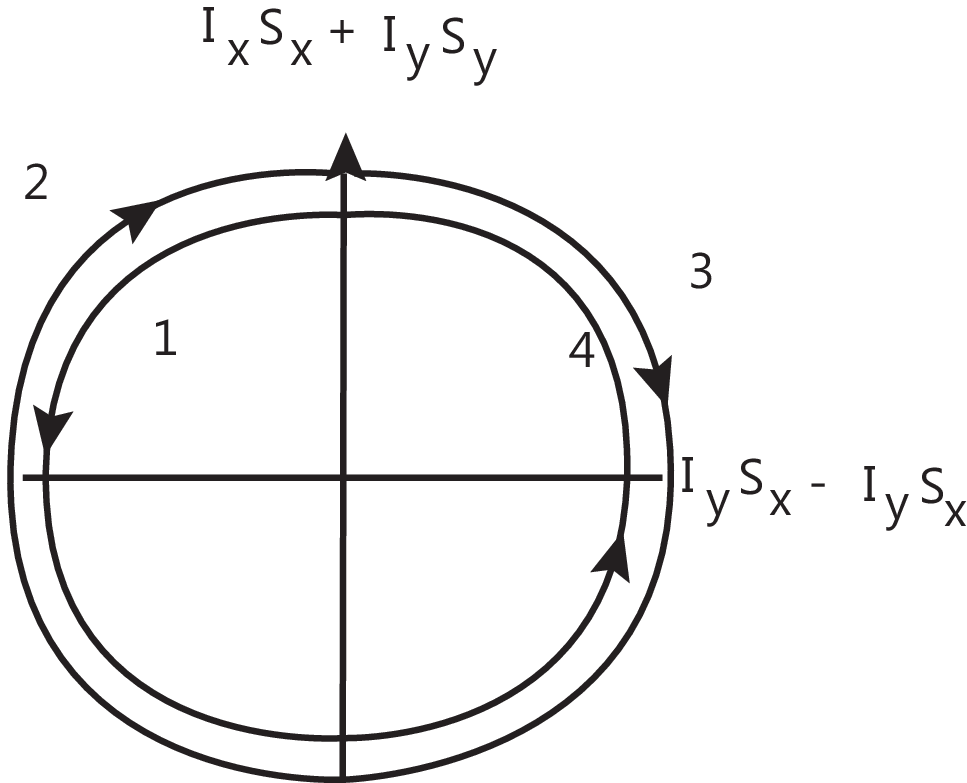}
\end{center}
\caption{ The top figure shows how planar part of the coupling evolves under four  stage pulse sequence in Eq. \ref{eq:largedip}. \label{fig:planarlarge}}
\end{figure}

Under large dip, the planar part of the coupling evolves in the toggling frame of the chemical shifts as shown in figure \ref{fig:planarlarge} and averages out.

The effective Hamiltonian in Eq. \ref{eq:efflargedip}, takes the following form, where $\theta = A \Delta$, 

\begin{eqnarray*}
H_{eff} &=& -\frac{\theta}{2} ( (\frac{\sin \frac{\omega_I \Delta}{2}}{\frac{\omega_I \Delta}{2}})^2 \omega_I I_y + (\frac{\sin \frac{\omega_S \Delta}{2}}{\frac{\omega_S \Delta}{2}})^2 \omega_S S_y ) +  \frac{\theta^2}{2} ((\frac{\sin \frac{\omega_I \Delta}{2}}{\frac{\omega_I \Delta}{2}})^2 (\frac{\sin \omega_I \Delta}{\omega_I \Delta}) \omega_I I_z \\ &+& (\frac{\sin \frac{\omega_S \Delta}{2}}{\frac{\omega_S \Delta}{2}})^2 (\frac{\sin \omega_S \Delta}{\omega_S \Delta}) \omega_S S_z ) + 2 \pi J \theta^2 ( \frac{1}{3} \frac{\sin \omega_I \Delta}{\omega_I \Delta} \frac{\sin \omega_S \Delta}{\omega_S \Delta} + f(\omega_I , \omega_S))I_yS_y.
\end{eqnarray*}where,


\begin{eqnarray*}
f(\omega_I , \omega_S) &=& \frac{1}{4 (\omega_S \Delta)^2}( \frac{\sin(\omega_I + \omega_S) \Delta}{(\omega_I + \omega_S)\Delta} +  \frac{\sin(\omega_I - \omega_S) \Delta}{(\omega_I - \omega_S)\Delta}) +  \frac{1}{4 (\omega_I \Delta)^2}( \frac{\sin(\omega_I + \omega_S) \Delta}{(\omega_I + \omega_S)\Delta} \\ &+&  \frac{\sin(\omega_S - \omega_I) \Delta}{(\omega_S - \omega_I)\Delta}) - \frac{1}{2 (\omega_S \Delta)^2}\frac{\sin \omega_I \Delta}{\omega_I \Delta} -  \frac{1}{2 (\omega_I \Delta)^2}\frac{\sin \omega_S \Delta}{\omega_S \Delta} + \frac{1}{6}\frac{\sin \omega_I \Delta}{\omega_I \Delta}\frac{\sin \omega_S \Delta}{\omega_S \Delta} 
\end{eqnarray*}

\begin{figure}[htb!]
\begin{center}
\includegraphics[scale=.5]{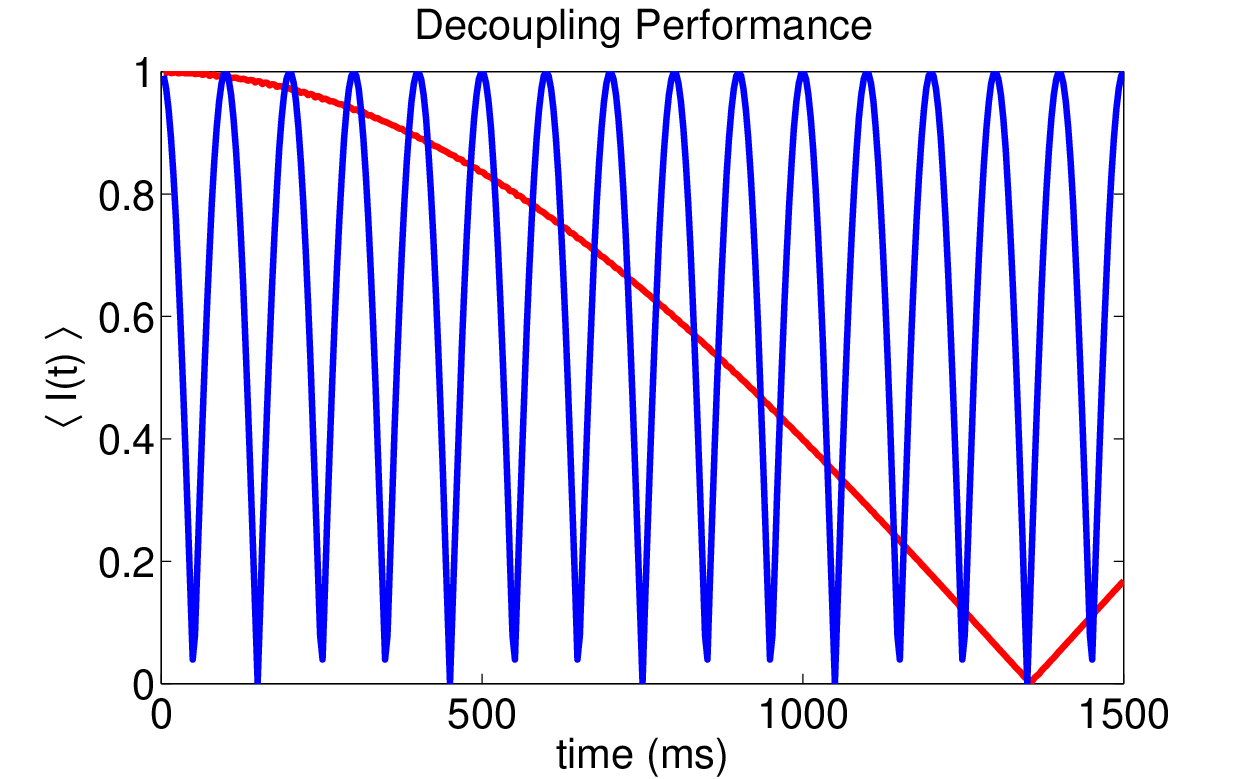}
\end{center}
\caption{ The top figure shows how magnetization on spin I evolves when we donot decouple (green curve) vs when we apply the four stage decoupling sequence in Fig. \ref{fig:hdpulsebasic} with large dip angle (red curve). The decoupling of spins results in magnetization of spin $I$ staying on spin $I$. Here, $\frac{\omega_I }{2\pi}= 2400$ Hz, $\frac{\omega_S}{2 \pi} = 800$ Hz, $J= 10$ Hz, $\frac{A }{2 \pi}= 200$ Hz and $\theta = \frac{\pi}{8}$} \label{fig:hdlargedip}
\end{figure}

Therefore a part of the coupling stays and reflects itself as an envelop in Fig. \ref{fig:hdlargedip}, where
we show how the magnetization on spin $I$ evolves when we donot decouple (blue curve) vs when we apply our decoupling sequence (red curve).
When no decoupling is performed, magnetization evolves to spin $S$ and we see a coupling evolution. When we apply decoupling sequence, the magnetization
on spin $I$ stays

In this paper we studied the problem of homonuclear decoupling. We first analyzed the case of Ising coupling and developed a four stage pulse sequence to decouple this interaction and measure chemical shifts. We then looked at Isotropic Hamiltonian and showed how the four stage sequence had to be modified to a six stage pulse sequence to decouple
isotropic coupling. We presented calculations and simulations to show the effectiveness of decoupling sequences.


\begin{thebibliography}{99}
\bibitem{ernst} R. Ernst, G. Bodenhausen, and A. Wokaun, Principles of Nuclear Magnetic
Resonance in One and Two Dimensions (Clarendon, 1987).

\bibitem{freeman} Freeman, Spin Choreography: Basic Steps in High Resolution NMR
(Oxford University Press, 1998).

\bibitem{shaka} A. J. Shaka, J. Keeler, and R. Freeman, ``Evaluation of a new broad-
band decoupling sequence: WALTZ-16,'' J. Mag. Reson. 53, 313-340
(1983).

\bibitem{starcuk} Z. Starcuk, K. Bartusek, and Z. Starcuk, ``Heteronuclear broadband spin-
flip decoupling with adiabatic pulses,'' J. Mag. Reson. 107, 24-31 (1994).

\bibitem{kupce} E. Kupce and R. Freeman, ``Adiabatic pulses for wideband inversion and
broadband decoupling, `` J. Mag. Reson. 115, 273-276 (1995).

\bibitem{cavanagh} J. Cavanagh, W. J. Fairbrother, A. G. PalmerIII, M. Rance, and N. J. Skel-
ton, Protein NMR Spectroscopy: Principles and Practices (Elsevier Academic, 2007).

\bibitem{mueller} M. A. McCoy and L. Mueller, ``Selective shaped pulse decoupling in NMR:
Homonuclear [carbon-13]carbonyl decoupling,''J. Am. Chem. Soc. 114,
2108-2112 (1992).

\bibitem{kupce} E. Kupce and G. Wagner, ``Multisite band-selective decoupling in proteins,''
J. Magn. Reson., Ser. B 110, 309-312 (1996).

\bibitem{wagner} E. Kupce and G. Wagner, ``Wideband homonuclear decoupling in protein
spectra,'' J. Magn. Reson., Ser. B 109, 329-333 (1995).

\bibitem{farjon} J. Farjon, W. Bermel, and C. Griesinger, ``Resolution enhancement in spec-
tra of natural products dissolved in weakly orienting media with the help
of 1 H homonuclear dipolar decoupling during acquisition: Application
to 1 H- 13 C dipolar couplings measurements,'' J. Mag. Reson. 180, 72-82
(2006).

\bibitem{goldman} M. Goldman, Quantum Description of High-Resolution NMR in Liquids
(Oxford University Press, 1988).

\bibitem{owrutsky} P. Owrutsky and N. Khaneja, ``Control of inhomogeneous ensembles on the
bloch sphere,'' Phys. Rev. A 86, 022315 (2012).

\end{thebibliography}
\end{document}